\documentclass[11pt,a4paper]{article}
\RequirePackage{amsmath,amssymb}
\RequirePackage[dvipsnames,usenames]{color}

\usepackage{soul}
\usepackage{cite}
\usepackage{fullpage}

\usepackage[british]{babel}
\usepackage[latin1]{inputenc}
\usepackage[T1]{fontenc}
\usepackage[final]{showkeys} 
\usepackage[bookmarks]{hyperref}
\usepackage{amsthm}
\usepackage{graphicx}
\usepackage{subfigure}
\usepackage{braket}
\usepackage{mathrsfs}
\usepackage{bm}

\newcommand{\be}{\begin{eqnarray}}
\newcommand{\ee}{\end{eqnarray}}
\def\eg{{\em e.g.}}
\def\ie{{\em i.e.}}

\renewcommand{\d}{\mbox{${\rm d}$}} 
\newcommand{\wt}[1]{\widetilde{#1}}

\newcommand{\lp}{\ell_{\rm P}}
\newcommand{\mpl}{m_{\rm P}}
\newcommand{\Gn}{G_{\rm N}}
\renewcommand{\a}{\hat a}
\newcommand{\ac}{\hat a^{\dagger}}

\title{\bf A quantum state for the late Universe}
\author{Andrea~Giusti$^{a}$\thanks{E-mail: agiusti@phys.ethz.ch},
$\ $
Silvia~Buffa$^{ab}$\thanks{E-mail: silvia.buffa@studio.unibo.it},
$\ $
Lavinia~Heisenberg$^{a}$\thanks{E-mail: laviniah@phys.ethz.ch}
$\ $
and 
Roberto~Casadio$^{bc}$\thanks{E-mail: casadio@bo.infn.it},
\\
\\
$^a$ {\em Institute for Theoretical Physics, ETH Zurich}
\\
{\em Wolfgang-Pauli-Strasse 27, 8093 Zurich, Switzerland}
\\
\\
$^b${\em Dipartimento di Fisica e Astronomia, Universit\`a di Bologna}
\\
{\em via Irnerio~46, 40126 Bologna, Italy}
\\
\\
$^c${\em I.N.F.N., Sezione di Bologna, I.S.~FLAG}
\\
{\em viale B.~Pichat~6/2, 40127 Bologna, Italy}
}
\begin{document}
\maketitle
%
%
\begin{abstract}
We consider the quantum description of a toy model universe in which the Schwarzschild-de~Sitter
geometry emerges from the coherent state of a massless scalar field.
Although highly idealised, this simple model allows us to find clear hints supporting the conclusion 
that the reaction of the de~Sitter background to the presence of matter sources induces i) a modified
Newtonian dynamics at galactic scales and ii) different values measured for the present Hubble
parameter.
Both effects stem from the conditions required to have a normalisable quantum state.
\par
\null
\par
\end{abstract}
%
%
%
%
%
%
\section{Introduction}
\label{sec:intro}
\setcounter{equation}{0}
General Relativity is the most successful theory of gravity and, within it, the $\Lambda$CDM model provides
the reference description of the late Universe, which thus appears dominated by a cosmological constant
$\Lambda$, a form of Dark Energy (DE), and (Cold) Dark Matter (DM).
One of the underlying assumptions in this picture is that the Universe at large scales can reliably be described
solely in terms of classical physics.
One might argue that all of physics should be described by quantum theories and that classical behaviours are
only approximately reproduced by suitable quantum states.
However, this perspective remains of purely academic interest unless a better understanding 
of observations is gained.
In this work, we will investigate the possibility that some of the features of the observed Universe, 
which appear as otherwise unexplained ingredients in the $\Lambda$CDM model, can indeed
be understood by considering a suitable quantum state for the DE and the gravitational field generated
by matter sources.
\par
For this purpose, we shall consider a highly idealised isotropic universe containing the cosmological constant
$\Lambda$ and just one localised matter source of Misner-Sharp-Hernandez mass
\be
M = 4\, \pi \int_0^{R_{\rm s}} r^2 \, \d r \, \rho (r)
\ ,
\ee
where $\rho=\rho(r)$ is the matter density profile and $R_{\rm s}$ the source radius.
In General Relativity, the corresponding geometry, in the exterior of the matter source, is given by the
Schwarzschild-de~Sitter (SdS) solution~\cite{FGB}
\be
\d s^2
=
-f(r)\,\d t^2
+
\frac{\d r^2}{f(r)}
+
r^2\,\d\Omega^2
\ ,
\label{eq:sds}
\ee
where $f=1+2\,V_{\rm SdS}$ and 
\be
\label{EqGeneralSolution}
V_{\rm SdS}
= - \frac{\Gn \, M}{r} - \frac{\Lambda \, r^2}{6}
\ .
\ee
This spacetime contains at most two horizons, determined by the roots of $f=0$.
In particular, for a compact source in a much larger universe, we can assume
$\sqrt{3/ \Lambda} \gg 2 \,\Gn\,M$ and one then finds the black hole horizon~\footnote{For a regular
matter source, $R_{\rm s}>R_{\rm H}$ and $R_{\rm H}$ in Eq.~\eqref{eq:Rh} is its gravitational radius.} 
\be
R_{\rm H} \approx 2 \,\Gn\, M
\label{eq:Rh}
\ee
and the much larger cosmological horizon
\be
L \approx \sqrt{3/ \Lambda}
\ .
\label{eq:LL}
\ee
We finally recall that $V_{\rm SdS}$ is the gravitational potential in the radial geodesic equation
\be
\frac{1}{2}\,m\left(\frac{\d r}{\d\tau}\right)^2
+
m\,V_{\rm SdS}
=
\frac{m}{2}
\left(
\frac{E}{m}
-1
\right)
\ ,
\ee
where $\tau$ is the proper time and $E=m\,f\,\d t/\d\tau$ is the conserved energy
of a test particle of proper mass $m$ moving in the region $R_{\rm H}\sim R_{\rm s}<r<L$. 
\par
Under the assumption of staticity and spherical symmetry, the Einstein--Hilbert action effectively
reduces to a scalar field theory in the weak-field limit~\cite{Casadio:2016zpl,Casadio:2017cdv}. 
Upon quantisation, one then finds that the classical gravitational potential in the exterior
of a compact source emerges as the mean-field approximation for a suitable coherent quantum state
of this effective scalar field~\cite{Casadio:2016zpl,Casadio:2017cdv} (and 
the classical geometry can be reconstructed accordingly~\cite{Casadio:2021eio,Casadio:2021gdf}).~\footnote{For a
similar description of the de~Sitter spacetime, see also Refs.~\cite{brahma}.}
In this perspective, the whole geometrical picture of gravity needs only emerge at the classical level
of the dynamics.
Moreover, the fact that the Einstein-Hilbert action suffers from the known issue of perturbative
non-renormalizability could be resolved if quantum states describing the (trans-)Planckian energy regime 
do not contain excitations for all of the modes which would build up divergences.
This (conjectured) phenomenon has been termed {\em classicalization\/} in the framework of
corpuscular gravity~\cite{classicalization,giusti} and a possible connection with asymptotic safety has also
been suggested~\cite{percacci}.
Some further indications for classicalization in the gravitational collapse can be found in
Ref.~\cite{Casadio:2021cbv}.
\par
Following the same line of reasoning, we shall assume that the SdS solution is described by a coherent state
$\ket{g}$ of a scalar field from which the classical potential $V_{\rm SdS}$ is recovered as a mean-field approximation
in the region $R_{\rm H}\sim R_{\rm s}\lesssim r\lesssim L$.
In the process of constructing the state $\ket{g}$, we shall see that the finite gravitational radius~\eqref{eq:Rh}
and cosmological horizon~\eqref{eq:LL} play a crucial role in ensuring that $\ket{g}$ is well-defined,
which leads to non-trivial consequences.
\par
In particular, we shall find that the coherent state $\ket{g}$ necessarily contains a contribution 
from ``soft scalar gravitons''~\footnote{These quanta are the analogue of virtual photons in the Coulomb
potential and can be used to describe any static classical field configuration~\cite{Mueck:2013mha,Bose:2021ytn}.}
corresponding to the {\em dark force\/} found in Refs.~\cite{Cadoni:2017evg,Cadoni:2018dnd,Cadoni:2020mgb},
which can reproduce the phenomenology at galactic scales usually attributed to the presence of DM.
In fact, we shall explicitly show the connection with Milgrom's Modified Newtonian Dynamics
(MOND)~\cite{Milgrom1} in Section~\ref{sec:MOND}.
Moreover, the presence of baryonic sources alters the cosmological horizon and this could help to 
alleviate (or even resolve) the tension between different measurements of the 
Hubble constant~\cite{DiValentino:2021izs}, as we shall discuss in Section~\ref{sec:H0}.
Even though the discrepancy could be just due to systematic errors, at this stage it cannot be discarded that it might signal
deviations from the $\Lambda$CDM model.
One possible resolution might arise from interactions between the DE and the DM sectors.
Such effective couplings in the dark sector can be motivated from stringy completion in the UV~\cite{Agrawal:2019dlm}.
An effective coupling between DE and DM can successfully give rise to a reduction of mass of DM while increasing the DE value,
leading to an increase in $H_0$.
If one can convert a sufficient amount of DM into DE, not only the $H_0$ tension can be reduced
but also the $\sigma_8$ tension since there would be less DM to form cosmic structures.
The same can be achieved in the presence of a  vector field with a phantom-like equation of state
parameter~\cite{Heisenberg:2020xak}.
In the present work, the changes in the quantum state of the system due to inclusion of matter to 
the de Sitter space will effectively give rise to an interacting DE-DM model in the low energy effective field theory. 
\par
Throughout this work we use units with $c=1$, but display explicitly
$\hbar=\lp\,\mpl$ and $\Gn=\lp/\mpl$, with $\lp$ the Planck length and $\mpl$ the Planck mass.
\section{The Universe as a macroscopic quantum system}
\label{sec:coherent}
\setcounter{equation}{0}
We start the quantum description of the idealised universe introduced in the previous section
by noticing that the potential $V_{\rm SdS}$ is an exact solution of
\be
\label{EqPoisson-Vacuum}
\triangle V = 4 \pi\, \Gn \, \rho - \Lambda
\ .
\ee
This Poisson-like equation in turn can be viewed as the static limit of a field theory defined
by the Lagrangian
\be
L [\Phi , J]
=
4\, \pi \int_0^\infty r^2\, \d r
\left( \frac{1}{2}\, \Phi \,\Box \Phi - J \, \Phi
\right)
\ ,  
\ee
for the canonically normalised scalar field $\Phi = {V(r)}/{\sqrt{\Gn}}$~\footnote{The
gravitational potential is dimensionless (in our choice of units), whereas a canonically normalised scalar field must have
dimensions of $($mass$/$length$)^{1/2}$, hence the factor of $\Gn^{-1/2}=\sqrt{{\mpl}/{\lp}}$~\cite{Casadio:2017cdv}.}
coupled to the current
\be
J
=
4\, \pi\, \sqrt{\Gn} \, \rho
-
\frac{\Lambda}{\sqrt{\Gn}}
\ .
\ee
We can then proceed with the canonical quantisation based on the normal modes of the corresponding
free field equation
\be
\Box
\Phi(t,r)
=
\left(-\partial_t^2+\triangle\right)
\Phi
=
0
\ .
\label{KG}
\ee
For our purposes, solutions to Eq.~\eqref{KG} can be conveniently written as~\cite{Casadio:2017cdv,Casadio:2021eio}
\be
u_{k}
=
e^{-i\,k\,t}\,j_0(k\,r)
\ ,
\ee
where $j_0(x)=(\sin x)/x$ is the spherical Bessel function of zero order.
For any function $F=F(r)$, we then have 
\be
\wt{F} (k)
=
4\,\pi \int _0 ^\infty r^2\, \d r \, j_{0} (k r) \, F(r) 
\ 
\Leftrightarrow
\ 
F (r)
=
\int _0 ^\infty \frac{k^2\, \d k}{2 \,\pi ^2}
\, j_{0} (k r) \, \wt{F} (k) 
\ ,
\label{eq:ft}
\ee
which allow us to write the quantum field operator in terms of creation and annihilation operators,
\be
\label{Phi}
\hat{\Phi}(t,r)
=
\int_0^\infty
\frac{k^2\,\d k}{2\,\pi^{2}}\,
\sqrt{\frac{\hbar}{2\,k}}
\left[
\a_{k}\,
u_k(t,r)
+
\ac_{k}\, 
u^*_k(t,r)
\right]
\ ,
\ee
acting on the Fock space built from the vacuum defined by $\a_{k}\ket{0}=0$ for all $k>0$.
\par
Given a solution $\Phi=\Phi _{\rm cl} (r)$ of Eq.~\eqref{EqPoisson-Vacuum}, one can always find a coherent
state $\ket{g_{\rm cl}}$ such that the classical solution is reproduced by the expectation value
\be
\bra{g_{\rm cl}} \hat{\Phi}(t,r) \ket{g_{\rm cl}} = \Phi _{\rm cl} (r)
\ .
\ee
The state $\ket{g_{\rm cl}}$ satisfies $\hat{a} _{k} \ket{g_{\rm cl}}=e^{i\, \gamma_{k}(t)}\, g_{\rm cl}(k) \ket{g_{\rm cl}}$,
with $\gamma_k=-k\,t$ and
\be
\label{eq:choerentstategeneral}
g_{\rm cl}(k) 
=
\sqrt{\frac{k}{2\, \hbar}}\, \wt{\Phi} _{\rm cl} (k)
\ .
\ee
Moreover, since
\be
\ket{g_{\rm cl}}
=
e^{-N_{\rm cl}/2}\,
\exp\left\{
\int_0^\infty
\frac{k^2\,\d k}{2\,\pi^2}\,
g_{\rm cl}(k)\,
\ac_k
\right\}
\ket{0}
\ ,
\label{gstate}
\ee
the total occupation number of ``scalar gravitons'' with respect to the vacuum $\ket{0}$ is given by 
\be
\label{eq:occupation}
N_{\rm cl}
=
\int\limits_{K_{\rm IR}}^{K_{\rm UV}}
\frac{k^2\, \d k}{2\, \pi ^2}\,|g_{\rm cl}(k)| ^2
\ ,
\ee
where we made explicit that the momentum $k$ for which $g_{\rm cl}(k)\not=0$ must in general range between a finite
minimum, encoded by the infrared (IR) cut-off ${K_{\rm IR}}$, and a finite maximum given by the ultraviolet (UV) cut-off
${K_{\rm UV}}$ in order to obtain a finite expression for $N_{\rm cl}$.
This condition is necessary for the corresponding $\ket{g_{\rm cl}}$ to be a normalisable, hence well-defined, state
and will play a crucial role in the following.
\par
Since the SdS background~\eqref{EqGeneralSolution} is an exact solution of Eq.~\eqref{EqPoisson-Vacuum},
we are interested in the case
\be
\label{eq:PhiCl-SdS}
\Phi _{\rm SdS}
=
\frac{V_{\rm SdS}}{\sqrt{\Gn}} = 
- \frac{\sqrt{\Gn} \, M}{r} - \frac{\Lambda \, r^2}{6\,\sqrt{\Gn}}
\ .
\ee
Taking the Fourier transform~\eqref{eq:ft} of Eq.~\eqref{eq:PhiCl-SdS}, one finds
\be
\wt{\Phi} _{\rm SdS} (k)
=
\wt{\Phi} _{M} (k) + \wt{\Phi} _{\Lambda} (k)
\ ,
\ee
with~\footnote{In asymptotically flat infinite space ($R_\infty\to\infty$), one can show that the first term in square brackets
does not contribute and $\wt{\Phi} _{M}$ yields the coherent state of Refs.~\cite{Casadio:2017cdv,Casadio:2021eio}.}
\be
\label{eq:PhiM}
\nonumber
\wt{\Phi} _{M} (k)
&\!\!=\!\!&
4 \,\pi\, \int_0^{R_\infty}
r^2\,\d r \, j_{0} (k r) 
\left( - \frac{\sqrt{\Gn} \, M}{r} \right)
\\
&\!\!=\!\!&
\frac{4\, \pi\, \sqrt{\Gn}\, M}{k^2}
\left[
\cos \left(k R_{\infty} \right)-1
\right]
\ee
and
\be
\label{eq:PhiLambda}
\nonumber
\wt{\Phi} _{\Lambda} (k) 
&\!\!=\!\!& 
4\, \pi\, \int_0 ^{R_\infty} 
r^2\d r \, j_{0} (k r)
\left(  - \frac{\Lambda \, r^2}{6\,\sqrt{\Gn}} \right)
\\
&\!\!=\!\!&
\frac{2 \,\pi\,  \Lambda}{3\, \sqrt{\Gn}\, k^5}  
\left[
k\, R_{\infty} \left(k^2\, R_{\infty}^2-6\right)
\cos\!\left(k\, R_{\infty}\right)
-
3 \left(k^2\, R_{\infty}^2-2\right)
\sin \left(k \,R_{\infty}\right)
\right]
\ .
\ee
where $0<R_{\infty}<\infty$ is an upper bound that is required to make both of the above expressions formally well-defined.
\par
From Eq.~\eqref{eq:choerentstategeneral}, the coherent state for the SdS solution is characterised by
\be
g_{\rm SdS}(k) = 
\sqrt{\frac{k}{2\, \hbar}}
\left[ \wt{\Phi} _{M} (k) + \wt{\Phi}_{\Lambda} (k) \right]
\ ,
\ee
which implies
\be
|g_{\rm SdS}(k)| ^2 = 
\frac{k}{2\, \hbar}
\left[ \wt{\Phi} _{M} ^2 (k) + \wt{\Phi} ^2 _{\Lambda} (k) 
+ 2\,\wt{\Phi} _{M} (k) \,\wt{\Phi} _{\Lambda} (k) \right]
\ .
\ee
Therefore, one can split Eq.~\eqref{eq:occupation} into three contributions, namely
\be
N_{\rm SdS} = N_{M} + N_{\Lambda} + N_{M\Lambda}
\ ,
\ee
with
\be 
N_{M} 
&\!\!=\!\!& 
\int\limits_{K_{\rm IR}} ^{K_{\rm UV}} 
\frac{k^3\, \d k}{4\, \pi ^2 \,\hbar}\,
\wt{\Phi} _{M} ^2 (k)
\simeq
\frac{\Gn \,M^2}{\hbar}\, \log\! \left(\frac{K_{{\rm UV}}}{K_{{\rm IR}}}\right)
\\
&&
\nonumber
\\
N_{\Lambda} 
&\!\!=\!\!& 
\int\limits_{K_{\rm IR}} ^{K_{\rm UV}} 
\frac{k^3 \, \d k}{4\, \pi ^2 \,\hbar}\,
\wt{\Phi} _{\Lambda} ^2 (k) 
\simeq 
\frac{\Lambda^2}{\Gn\, \hbar\,  K_{\text{UV}}^6}
\left[
\left(\frac{K_{{\rm UV}}}{K_{{\rm IR}}}\right)^6
\log\! \left(\frac{K_{{\rm UV}}}{K_{{\rm IR}}}
\right)
\right]
\ee
and
\be
N_{M\Lambda} 
&\!\!=\!\!& 
\int\limits_{K_{\rm IR}} ^{K_{\rm UV}} 
\frac{k^3 \, \d k}{4\, \pi ^2 \,\hbar}\,
\wt{\Phi} _{M} (k)\, \wt{\Phi} _{\Lambda} (k) 
\simeq
\frac{\Lambda \,M}{\hbar\,  K_{\text{UV}}^3} 
\left[
\left(\frac{K_{{\rm UV}}}{K_{{\rm IR}}}\right)^3
\log\! \left(\frac{K_{{\rm UV}}}{K_{{\rm IR}}}\right)
\right]
\ ,
\ee
where the final expressions are the leading order terms in $K_{{\rm UV}}/K_{{\rm IR}}\gg 1$ and
$\simeq$ denotes equality up to order-one multiplicative factors.
As we mentioned above, the momentum integration is effectively restricted to a finite range by 
$\wt{\Phi} _{\rm SdS}(k<K_{\rm IR})=\wt{\Phi} _{\rm SdS}(k>K_{\rm UV})=0$, otherwise
$N_{\rm SdS}$ would not be finite and $\ket{g_{\rm SdS}}$ would be ill-defined.
Furthermore, since the integration range in Eqs.~\eqref{eq:PhiM} and~\eqref{eq:PhiLambda} 
is limited above by $R_{\infty}$, the momenta $0<k < 1/R_{\infty}\sim K_{\rm IR}$ must necessarily
be excluded. 
\par
So far the IR and UV cut-offs have appeared as formal requirements, but we can now give them a clear physical meaning.
As we recalled in the Introduction, the SdS spacetime of interest to us is contained within the two length
scales $R_{\rm H}$ in Eq.~\eqref{eq:Rh} and $L$ in Eq.~\eqref{eq:LL}.
The mean-field description does not need to reproduce the classical behaviour in regions 
beyond these two horizons, hence the coherent state can satisfy the weaker condition~\footnote{For $L\to\infty$, the
effect of the cosmological constant disappears and one obtains the results of Ref.~\cite{Casadio:2021eio}.}
\be
\bra{g} \hat{\Phi}(t,r) \ket{g} 
\simeq
\frac{V_{\rm SdS}(r)}{\sqrt{\Gn}}
\qquad 
\mbox{for}
\ \
R_{\rm H} \lesssim r \lesssim L
\ , 
\ee
where the equality needs only to hold within experimental bounds.
Since modes with $k\ll1/L$ do not contribute significantly for a system of size $L$, we can set the IR cut-off 
$K_{\rm IR}\sim 1/L$. 
The UV cut-off is similarly determined by $K_{\rm UV}\sim 1/R_{\rm H}$, since the SdS solution does not apply inside the
compact source of radius $R_{\rm s}\gtrsim R_{\rm H}$.
By employing these values for the cut-offs, we find
\be
N_{M} 
&\!\!\simeq\!\!&  
\frac{M^2}{\mpl ^2}\, \log\!\left(\frac{\mpl \, L}{M \, \lp}\right)
\label{eq:NM}
\\
&&
\nonumber
\\ 
N_{\Lambda} 
&\!\!\simeq\!\!& 
\frac{L^2}{\lp^2}\, \log\! \left(\frac{\mpl \, L}{M \, \lp}\right)
\label{eq:NL}
\ee
and
\be
N_{{M}\Lambda} 
&\!\!\simeq\!\!&
\frac{M \, L}{\mpl\, \lp}\, \log\! \left(\frac{\mpl \, L}{M \, \lp}\right)
\ .
\label{eq:NML}
\ee
where we used $\hbar = \mpl \,\lp$, $\Gn = \lp/\mpl$, $L \simeq 1/\sqrt{\Lambda}$, and $R_{\rm H} \simeq \Gn \,M$.
\par
The total occupation number of the coherent state $\ket{g}$ for our toy universe finally reads
\be
\label{Ntotal}
N
\simeq
\left(
\frac{M^2}{\mpl ^2} + \frac{L^2}{\lp^2} + \frac{M  L}{\mpl\, \lp}
\right) 
\log\left(\frac{\mpl \, L}{M \, \lp}\right)
\ .
\ee
Note that $N_{{M}}$ and $N_{\Lambda}$ exactly match the scaling laws resulting
from the corpuscular description of the Schwarzschild black hole and the de~Sitter space,
respectively (up to the logarithmic factor, see Refs.~\cite{DvaliGomez,Dvali:2013eja,giusti}).
While the complete expression for $N_{{\rm M}}$ was first derived in Ref.~\cite{Casadio:2017cdv},
the computation presented here provides the first equivalent derivation of the scaling law of $N_\Lambda$ 
for the de~Sitter space.
\section{Modified Newtonian dynamics}
\label{sec:MOND}
\setcounter{equation}{0}
The additional $N_{{M}\Lambda}$ in Eq.~\eqref{eq:NML} enters as a typically quantum cross-term between
the component $\wt{\Phi} _{\Lambda}$ accounting for the de~Sitter global geometry and $\wt{\Phi} _{M}$
reproducing the local gravitational interaction.
Furthermore, this additional contribution matches the predictions of corpuscular gravity for the emergence
of an effective {\em dark force\/} responsible for the modification of Newtonian physics at galactic
scales~\cite{Cadoni:2018dnd}.
These results hint at a theoretical explanation for DM phenomenology that does not require
the {\em ad hoc\/} addition of exotic matter.  
\par
A well-known alternative to the DM paradigm of modern cosmology is provided by Milgrom's
MOND~\cite{Milgrom1,Milgrom4}.
This effective non-relativistic and low-energy empirical model for gravity modifies Newton's law
when the acceleration of test particles falls below a critical scale $a_0$.
This modification then provides a fairly accurate description of the flattening of galaxy
rotation curves (see, \eg,~Ref.~\cite{Brouwer:2021nsr}). 
In order to see the connection with the model universe of Section~\ref{sec:coherent}, we first note that the critical
acceleration scale $a_0$ is equivalent to a critical length scale $\ell\equiv\sqrt{\Gn\, M / a_0}$, where $M$ is
the mass of a galaxy.
Astronomical observations set $a_0 \simeq H_0$, with $H_0$ denoting the Hubble
parameter today,~\footnote{For very recent analyses supporting this result,
see Refs.~\cite{Milgrom:2020cch,mcgaugh}.}
which implies $\ell = \sqrt{\Gn\, M / H_0}$.
Assuming that the late Universe is effectively described by the de~Sitter space at very large scales, 
$L = 1/H_0$ defines the cosmological horizon, and we find 
\be
\ell = \sqrt{\Gn\, M\, L}
\ .
\ee
In other words, the critical scale of MOND is determined precisely by the two fundamental scales,
$R_{\rm H}\simeq \Gn M$ and $L$, of the toy universe described in the previous section.
\par
MOND predicts that the effective gravitational potential in the outermost regions of a large distribution
of matter, like a galaxy, should follow the logarithmic scaling (see, \eg, Refs.~\cite{Cadoni:2017evg,Giusti:2020rul})
\be
V _{\rm MOND}
\simeq
\frac{\Gn\, M}{\ell} \, \log\!\left( \frac{r}{\ell} \right)
\ .
\label{Vmond}
\ee
Assuming this potential as our classical field configuration, we can construct the corresponding coherent state
by repeating the steps in Section~\ref{sec:coherent}, which yield
\be
g_{\rm MOND}(k)
=
\sqrt{\frac{k}{2\, \hbar}} \, \frac{\wt{V} _{\rm MOND}(k)}{\sqrt{\Gn}}
\simeq
-\frac{\Gn\, M}{\lp \, \ell\,k^{5/2}}
\ ,
\ee
where $\simeq$ denotes that two quantities are equal up to order-one multiplicative factors and
the necessary cut-offs have been included.
The corresponding occupation number reads
\be
N_{\rm MOND}
&\!\!=\!\!&
\int\limits_{K_{\rm IR}}^\infty 
\frac{k^2\,\d k}{2\, \pi ^2}\,|g_{\rm MOND}(k)| ^2
\simeq
\frac{\Gn^2\, M^2\, R_{\infty}^2}{\lp^2 \, \ell ^2}
\nonumber
\\ 
&\!\!\simeq\!\!&
\frac{\Gn\, M\, R_{\infty}^2}{\lp^2 \, L}
\ ,
\ee
where $K_{\rm IR} = 1/R_{\infty}$.
If we embed this scenario within a universe filled with vacuum energy described by the cosmological constant $\Lambda$,
we have the natural cut-off $K_{\rm IR}\sim 1/R_{\infty} \sim 1/L$ and
\be
\label{eq:Nmond}
N_{\rm MOND}
\simeq
\frac{M\, L}{\mpl\, \lp} 
\equiv 
N_{\rm DF}
\ ,
\ee
which reproduces the prediction based on a simple energy balance in Refs.~\cite{Cadoni:2017evg, Cadoni:2018dnd}.  
\par
Comparing Eqs.~\eqref{eq:NML} and \eqref{eq:Nmond}, we observe that, despite the absence of an explicit MOND
contribution~\eqref{Vmond} in the classical SdS potential~\eqref{EqGeneralSolution}, the quantum coherent state for the
SdS system obtained in Section~\ref{sec:coherent} contains the additional contribution of $N_{M\Lambda} \simeq N_{\rm DF}$
(up to logarithmic factors) scalar gravitons.
We remark once more that this result explicitly follows once the necessary UV and IR cut-offs are included to make
sense of the quantum picture and supports the conclusion that MOND-like effects naturally emerge
as part of the response of the collective DE state $\wt{\Phi} _{\Lambda}$ to the local presence of (baryonic)
matter~\cite{Cadoni:2017evg,Cadoni:2018dnd}.
\section{Implication for the $H_0$ tension}
\label{sec:H0}
\setcounter{equation}{0}
So far, we have discarded any consequences of the baryonic matter on the actual size of the cosmological horizon. 
In order to estimate such effects, let us consider an empty de~Sitter spacetime with horizon $\bar{L}=\sqrt{3/\Lambda}$.
The occupation number of the corresponding coherent state would be given by~\cite{Dvali:2013eja,Casadio:2020nns}
\be
\label{Ndec}
\bar N
\simeq
\frac{\bar{L}^2}{\lp^2}
\ .
\ee
If one now adds a baryonic source, \eg, a galaxy of mass $M$, the quantum state will readjust to account for the local
gravity and the competition between the local and cosmological holographic scalings will induce the emergence
of a dark force reproducing MOND at galactic scales~\cite{Cadoni:2017evg,Cadoni:2018dnd}, as we have seen
in the previous section.
\par
In particular, once we add the localised baryonic source, the total occupation number of the coherent state should be given 
according to Eq.~\eqref{Ntotal} by
\be
\label{Nlate}
N
\simeq
\frac{M^2}{\mpl ^2} + \frac{L^2}{\lp^2} + \frac{\Gn M  L}{\lp^2}
\ ,
\ee
with $L$ denoting the new size of the cosmological horizon (and we discarded logarithmic factors for simplicity).
If we assume that the addition of matter does not change the total number of scalar gravitons, but only their
distribution $g(k)$ in momentum space, equating $\bar N$ in Eq~\eqref{Ndec} to $N$ in Eq.~\eqref{Nlate} yields
\be
\label{eq:L}
\frac{\bar{L}^2}{\lp^2}
\simeq
\frac{M^2}{\mpl ^2} 
+ \frac{L^2}{\lp^2} 
+ \frac{\Gn\, M\,  L}{\lp^2}
\ ,
\ee
from which we can immediately infer that $\bar{L} > L$.
Then, solving Eq.~\eqref{eq:L} for $L$ and expanding the result to $\mathcal{O} (N_{\rm M}^3/N_{\rm \Lambda}^3)$,
we find~\cite{Cadoni:2018dnd}
\be
\label{eq:LbarL}
L 
\simeq 
\bar{L}
-
\frac{\lp\,M}{2 \, \mpl}
\left(1  + \frac{3\,M \, \lp}{4\,\mpl\, \bar{L}}
\right)
\ .
\ee
This result can be used to address different measurements of the present Hubble constant $H_0$.
\par
The physics of the early Universe should be related to a quasi-de~Sitter configuration, thus it is reasonable
to assume that $\bar{N}$ should describe this early phase and that the value of $H_0$ obtained
from the Cosmic Microwave Background (CMB) radiation should carry information about $\bar{L}$.
Hence, we simply assume that $H_0^{\rm CMB} \simeq 1/\bar{L}$.
On the other hand, $L$ is related to the late time evolution of the Universe, thus measurements of $H_0$
from Type~Ia supernovae should carry information about the latter, \ie, $H_0 ^{\rm SNeIa} \simeq 1/L$.
We already have a rough consistency check in this respect, since Eq.~\eqref{eq:LbarL} tells us that $L<\bar{L}$,
which implies that $H_0 ^{\rm SNeIa} > H_0 ^{\rm CMB}$.
\par
In order to refine our estimate, instead of a single clump of matter, let us consider $n_{\rm g} \gg 1$ galaxies
of typical mass $M_{\rm g}$ distributed over the de~Sitter configuration of our macroscopic quantum state.
Neglecting non-linear contributions among galaxies, we can generalise Eq.~\eqref{eq:L} to $n_{\rm g}$
sources as
\be
\bar{L}^2
&\!\!\simeq\!\!&
L^2 
+ 
n_{\rm g}\,\lp^2
\left(
\frac{M_{\rm g}^2}{\mpl^2} + \frac{\Gn\, M_{\rm g} \, L}{\lp^2}
\right)
\nonumber
\\
&\!\!\simeq\!\!&
L^2 + \Gn\, M_{\rm tot}\, L 
+ \frac{\Gn^2\, M_{\rm tot}^2}{n_g}
\ ,
\ee
where $M_{\rm tot} = n_{\rm g} \, M_{\rm g}$.
We then have
\be
\label{LastLbar}
\bar{L}
\simeq
L
\left[
1
+
\frac{\Gn\, M_{\rm tot}}{L}
\left(
1
+
\frac{\Gn\, M_{\rm tot}}{n_{\rm g}\,L}
\right)
\right]
\ .
\ee
\par
If we assume a $5\%$ component of baryonic matter in the Universe, 
we can insert $\Gn\, M_{\rm tot} \simeq 0.05 \, L$ into Eq.~\eqref{LastLbar} and obtain
\be
\bar{L} \simeq L \left( 1.05 + \frac{0.0025}{n_g} \right)
\ ,
\ee
which implies
\be
\frac{H_0 ^{\rm SNeIa} - H_0 ^{\rm CMB}}{H_0 ^{\rm CMB}}
\simeq 
0.05
+
\mathcal{O} \left( \frac{1}{n_g} \right)
\, ,
\ee
suggesting a $5\%$ discrepancy between $H_0 ^{\rm SNeIa}$ and $H_0 ^{\rm CMB}$.
This estimate comes surprisingly close to the observed $H_0$ tension~\cite{DiValentino:2021izs},
particularly considering that the actual time evolution of the late Universe was not accounted for at all 
in our expression of $H_0 ^{\rm CMB}$. In our formalism, the discrepancy of the Hubble parameter $H_0$ 
between the early Universe and the late Universe would be an expected phenomena and due to the effects of the
baryonic sources to the occupation number of the coherent states. This represents a completely new
and quantum resolution compared to the existing ones in the literature, which typically either rely on the presence of
an additional Quintessence/Horndeski field or a Generalized Proca field or an effective coupling in the
dark sector \cite{DiValentino:2021izs,Agrawal:2019dlm,Heisenberg:2020xak,Banerjee:2020xcn}. 
\section{Conclusions and outlook}
\label{sec:conclusions}
In this work we developed the quantum description of a toy model for the universe
containing only the cosmological constant and a localised matter source.
Despite its simplicity, this model allowed us to reach some interesting conclusions. 
First, we formally derived the corpuscular scaling law for the de~Sitter spacetime
by means of a purely (though simplified) field-theoretic approach.
Second, we showed that the competition between the two components of the quantum state,
respectively responsible for the cosmic expansion and for the local gravitational interaction,
gives rise to an effective dark force, akin to the one derived from corpuscular gravity in
Refs.~\cite{Cadoni:2017evg,Cadoni:2018dnd}, 
which reproduces the MOND phenomenology at galactic scales.
Finally, we noted that the addition of baryonic matter to the de~Sitter configuration modifies
the quantum state of the system in a way that also appears to relieve the observed $H_0$ tension. 
\par
Among the limitations of the model, overall staticity and the simplified description of 
each galaxy (and of the population of galaxies in the universe) are the most relevant.
Clearly, these assumptions provide a workable oversimplification of the problem, 
but they will have to be thoroughly reconsidered when trying to extend the analysis
to more realistic (and testable) scenarios.
\par
In particular, we can foresee at least two lines of further development.
First of all, in the present work we have only estimated the effects due to the coherent
state containing only momenta in a finite range, namely
$\wt{\Phi} _{\rm SdS}(k\lesssim 1/L)=\wt{\Phi} _{\rm SdS}(k>1/R_{\rm H})=0$.
For a more quantitatively analysis, the IR cut-off $K_{\rm IR}\sim 1/L$ and the UV
cut-off $K_{\rm UV}\sim 1/R_{\rm H}$ will have to be determined by a refined 
description of the observable de~Sitter patch and matter sources, respectively. 
An effective spacetime metric will then have to be reconstructed 
by replacing $V_{\rm SdS}$ in Eq.~\eqref{eq:sds} with 
\be
V_{\rm Q}(r)
=
\int\limits_{K_{\rm IR}}^{K_{\rm UV}}
\frac{k^2\, \d k}{2 \,\pi ^2}
\, j_{0} (k r) \, \wt{\Phi} _{\rm SdS} (k)
\ .
\label{eq:Vq}
\ee
This will allow us to study geodesics and place detailed bounds on $K_{\rm IR}$
and $K_{\rm UV}$ from experimental data. 
Moreover, an effective metric 
description would serve as a fundamental tool to further analyze the $H_0$ tension as well as 
investigate the implications of this model for the $\sigma_8$ anomaly. Typically, resolutions of the 
Hubble tension worsen the $\sigma_8$ tension. A general DE model needs to satisfy very specific requirements 
in order to perfectly balance the increase in the expansion with a decrease in the structure formation, which is
being investigated somewhere else. Successful examples are interacting DE-DM models and late Universe
cosmologies based on a vector field.
A second line was already mentioned in Section~\ref{sec:H0} and consists in extending
the construction of the quantum state of the universe to include multiple matter sources,
possibly including some non-linear contributions due to the local gravitational interactions.
Both lines will require us to rely (heavily) on numerical methods, since most expressions, like 
Eq.~\eqref{eq:Vq}, cannot be computed analytically and the number of galaxies in the
Universe is very large.
\section*{Acknowledgments}
A.G.~is supported by the European Union's 
Horizon 2020 research and innovation programme under the Marie Sk\l{}odowska-Curie Actions (grant 
agreement No.~895648--CosmoDEC). 
S.B.~is partially supported by a University of Bologna fellowship for master students.
R.C.~is partially supported by the INFN grant FLAG.
L.H.~is supported by funding from the European Research Council (ERC) under the European
Union's Horizon 2020 research and innovation programme grant agreement No.~801781 
and by the Swiss National Science Foundation grant No.~179740.
The work of R.C.~and A.G~has also been carried out in the framework of activities of the Italian National
Group of Mathematical Physics  [Gruppo Nazionale per la Fisica Matematica (GNFM), Istituto 
Nazionale di Alta Matematica (INdAM)]. 
\end{document}